\documentclass{article}

\usepackage[accepted]{icml2025}
\usepackage[most]{tcolorbox}
\usepackage{microtype}
\usepackage{graphicx}
\usepackage{subfigure}
\usepackage{booktabs} 

\usepackage{hyperref}

\usepackage[accepted]{icml2025}

\usepackage{amsmath}
\usepackage{amssymb}
\usepackage{mathtools}
\usepackage{amsthm}

\usepackage[capitalize,noabbrev]{cleveref}

\theoremstyle{plain}
\newtheorem{theorem}{Theorem}[section]

\theoremstyle{definition}

\theoremstyle{remark}

\usepackage[textsize=tiny]{todonotes}

\icmltitlerunning{}

\newcommand{\fakeparagraph}[1]{\noindent\textbf{#1.}}

\usepackage{xspace} 

\newcommand{\eg}{{\it e.g.,}\xspace}

\newcommand\figref[1]{Fig.~\ref{#1}}

\newcommand\tabref[1]{Table~\ref{#1}}

\newcommand\secref[1]{\S\ref{#1}}

\newcommand\appref[1]{Appendix~\ref{#1}}
\newcommand\algref[1]{Alg.~\ref{#1}}

\usepackage{nicematrix}
\def \tool{\texttt{DyCodeEval}\xspace}
\def \metric{\texttt{DyPass}\xspace}
 \usepackage{multirow}

\usepackage{xcolor}
\definecolor{pycomment}{RGB}{106,153,85}
\definecolor{pyblue}{RGB}{50,150,250}
\definecolor{pyred}{RGB}{250,80,80}
\definecolor{pygreen}{RGB}{100,200,100}

\tcbuselibrary{listingsutf8}

\icmltitlerunning{Dynamic Benchmarking of Code LLMs}

\begin{document}

\twocolumn[
\icmltitle{\tool: Dynamic Benchmarking of Reasoning Capabilities \\ in Code Large Language Models Under Data Contamination}




\begin{icmlauthorlist}
\icmlauthor{Simin Chen}{yyy}
\icmlauthor{Pranav Pusarla}{yyy}
\icmlauthor{Baishakhi Ray}{yyy} 

\end{icmlauthorlist}

\begin{center}
\large $^{1}$Columbia University
\end{center}

\icmlaffiliation{yyy}{Department of Computer Science, Columbia University}

\icmlcorrespondingauthor{Simin Chen}{sc5687@columbia.edu}


\vskip 0.3in
]




\printAffiliationsAndNotice{}  

\begin{abstract}

The rapid advancement of code large language models (Code LLMs) underscores the critical need for effective and transparent benchmarking methods. However, current benchmarking predominantly relies on publicly available, human-created datasets. The widespread use of these static benchmark datasets makes the evaluation process particularly susceptible to data contamination—an unavoidable consequence of the extensive data collection processes employed during LLM training.
Existing methods for addressing data contamination typically face significant limitations, including reliance on substantial human effort and difficulty in managing class imbalances. To overcome these challenges, we propose \tool, a novel benchmarking suite specifically designed to evaluate Code LLMs under realistic contamination scenarios. Given an initial seed programming problem, \tool utilizes multiple agents to systematically extract and modify contextual information without changing the core logic, generating semantically equivalent variations.
We introduce a dynamic data generation method and conduct extensive empirical studies on two seed datasets involving 18 Code LLMs. The results demonstrate that \tool effectively assesses the reasoning capabilities of Code LLMs under contamination conditions while producing diverse problem variants, thereby ensuring robust and consistent benchmarking outcomes. 
Our project webpage can be found at this link\footnote{\url{https://codekaleidoscope.github.io/dycodeeval.html}}.

\end{abstract}

\section{Introduction}


Large language models (LLMs) have demonstrated significant potential as assistant software developers, particularly in code generation~\cite{humaneval, guo2024deepseek, jiang2024self, di2024codefuse}. Consequently, numerous code-focused LLMs have been developed. These models are trained on vast corpora of natural language and programming language data. Once well trained, they can comprehend human instructions and generate the corresponding code snippets. 



As diverse model architectures and training algorithms for code LLMs continue to emerge~\cite{vaswani2017attention, shazeer2017outrageously}, a key focus in code LLM research is the effective benchmarking of each model's code reasoning capability. 
Without a standardized and transparent benchmarking suite, assessing these models' performance and driving improvements becomes a significant challenge.


However, existing benchmarking suites for evaluating code LLMs are inadequate due to their static benchmarking schema, which can lead to potential data contamination from unintended data crawling. Research suggests that such contamination may already be present in current LLMs~\cite{chen2025recent, jain2024livecodebench, dong2024generalization}.  
Although some methods aim to provide contamination-free benchmarking for code LLMs, they still rely on manual efforts. For example, \texttt{LiveCodeBench} \cite{jain2024livecodebench} proposes crawling new programming problems from online platforms and benchmarking LLMs based on timestamps, while \texttt{PPM} \cite{chen2024ppm} attempts to systematize new programming problems by combining manually defined operators.
However, these methods have several limitations: (1) \textit{Significant Manual Effort}: These methods still require substantial manual input to create such datasets. For example, \texttt{PPM} necessitates manually defining the lambda operator, while \texttt{LiveCodeBench} shifts the burden of manual design to question authors on coding platforms. (2) \textit{Imbalanced Semantic Complexity}: The newly generated benchmarking datasets often lack semantic equivalence with the original ones. As a result, when a model performs worse on these benchmarks, it is challenging to determine whether the lower score reflects diminished model capabilities or increased benchmark complexity. Thus, these new benchmark results fail to provide meaningful guidance for model developers to improve their models effectively.

To address this limitation, rather than manually creating benchmarking datasets with uncertain semantic complexity, we aim to develop an automated method for dynamically evaluating code LLMs. However, designing such a method presents two key challenges: (1)  \textit{Generating Semantically Diverse Yet Complexity-Controlled Problems}. The first challenge is how to ensure the generated problems vary in semantics while maintaining controlled complexity. (2) \textit{Providing Comprehensive Benchmarking}. A proper benchmark programming problem must include fine-grained test cases and canonical solutions to rigorously assess correctness.

To address these challenges, we draw inspiration from metamorphic testing \cite{chen2018metamorphic}, a widely used approach in software testing to tackle the oracle problem. In our case, we leverage the principles of metamorphic testing to automate comprehensive benchmarking.  
Specifically, we define a metamorphic relationship for programming problems.  A programming problem includes \textit{complexity-related algorithmic abstraction} and \textit{complexity-unrelated context description}. Modifying the \textit{complexity-unrelated context description} alters the problem’s semantics without changing its inherent complexity.   Building on this relationship, \tool employs LLM-based agents to generate diverse contexts for a seed problem, automatically transforming existing problems into semantically varied yet complexity-preserving versions. Additionally, \tool integrates a validation agent as a probabilistic oracle to verify the correctness and consistency of the newly generated problems, ensuring reliability.

We used \tool to generate new evaluation sets to assess Code LLM performance under both data contamination and real-world benchmarking scenarios. Our key findings are as follows:  
\begin{enumerate}
    \item Our method effectively reflects Code LLMs' reasoning capabilities in a manually crafted contamination environment (\secref{sec:contaminated}).  

    \item The performance of some Code LLMs on our dynamic benchmarks degraded significantly, suggesting potential data contamination of these Code LLMs (\secref{sec:wild}).  

    \item \tool generates semantically diverse programming problems, and its inherent randomness makes the likelihood of generating identical problems extremely low, thereby reducing the risk of data contamination (\secref{sec:diversity}).  
    
    \item Despite its randomness, \tool consistently produces stable benchmarking results, ensuring reliable evaluation (\secref{sec:stable}).
\end{enumerate}

We summarize our contribution as follows:
\begin{itemize}
    \item \textbf{Novel Problem Characterization.} We identify a limitation in current static benchmarking schemas, as they are insufficient for effectively evaluating modern Code LLMs, especially when data contamination occurs and the model’s training process lacks transparency.

    \item \textbf{New Methodology Design}. We propose a novel approach that separates \textit{context} and \textit{algorithm} in programming problems. Building on this concept, we introduce a dynamic benchmarking method, \tool, which generates programming problems for benchmarking without introducing additional complexity to the dataset. This approach mitigates the impact of data contamination, ensuring transparent and reliable benchmarking.

    \item \textbf{Empirical Findings}. We conduct an empirical evaluation of \tool, and the results demonstrate that traditional static benchmarks can create a false sense of accuracy. In contrast, our dynamic benchmarking approach provides consistently reliable results, even under data contamination scenarios. Additionally, \tool generates semantically diverse programming problems while maintaining stable benchmarking results.
    
\end{itemize}


\begin{figure*}
    \centering
    \includegraphics[width=0.9\textwidth]{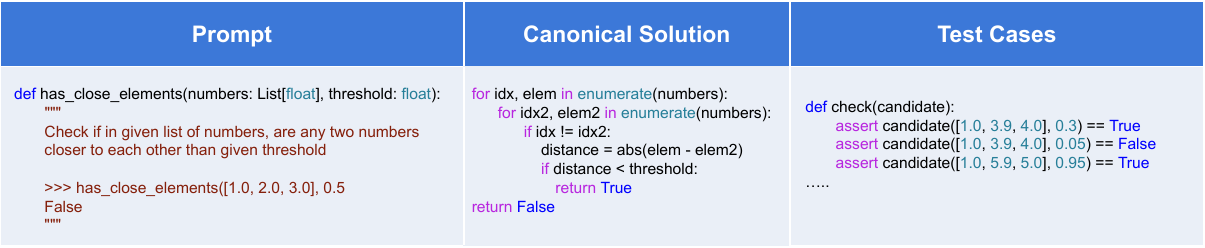}
    \caption{Benchmark programming problem example}
    \label{fig:humaneval}
\end{figure*}

\section{Background \& Related Work}



\subsection{Benchmarking Code LLMs}
\label{sec:benchmark_bg}

Code LLMs have been widely adopted in various real-world software engineering applications, leading to the development of numerous benchmarks for evaluating their capabilities in code understanding and reasoning~\cite{humaneval, li2024evocodebench, guan2025your, mbpp, chen2024ppm, mbpp, yu2024codereval, jimenezswe, ding2023crosscodeeval, mathai2024kgym}. Among the many tasks designed to assess code reasoning, this work focuses specifically on the task of natural language to code generation and reviews representative benchmarks in this area.
HumanEval~\cite{humaneval} introduced a human-crafted dataset to evaluate the code generation capabilities of large language models. EvalPlus~\cite{evalplus} later identified the limitations of HumanEval and MBPP—particularly their limited test case coverage—and proposed a more rigorous benchmark. HumanEval-XL~\cite{peng2024humaneval} further extended HumanEval to support multilingual settings.
\figref{fig:humaneval} illustrates an example from HumanEval, a widely used benchmark for the natural language to code generation task. Each programming problem typically consists of three components: a prompt, a canonical solution, and a set of test cases. The prompt is first fed into the Code LLM to generate a candidate solution, which is then executed against hidden test cases to evaluate its correctness.

\subsection{Data Contamination Free Benchmarking}

Data contamination has become a significant concern in benchmarking large language models (LLMs)~\cite{languagemodelsarefewshotlearners, jain2024livecodebench, chen2025recent}, as it can lead to inflated performance scores and unreliable evaluations. To mitigate this issue, researchers have proposed various contamination-free benchmarking strategies, which can be broadly categorized into three approaches.
The first line of work focuses on data protection through encryption and privatization. For instance, Jacovi et al.\cite{jacovi-etal-2023-stop} and Rajore et al.\cite{rajore2024truce} propose techniques to safeguard benchmark data from being included in LLM training corpora.
The second line of research emphasizes timely benchmark updates. LiveBench~\cite{white2024livebench}, for example, compiles questions from recent sources such as math competitions held within the past year and regularly updates its dataset. Similarly, LiveCodeBench~\cite{jain2024livecodebench} continuously collects new human-authored programming problems from online platforms like LeetCode to maintain freshness and reduce the risk of contamination.
The third line of research explores dynamic generation of evaluation sets. DyVal~\cite{zhu2024dyval} uses DAG structures to create dynamic benchmarks, TreeEval~\cite{li2024treeeval} employs high-performing LLMs to generate and evaluate problems via tree planning, and ITD (Inference-Time Decontamination)~\cite{zhu-etal-2024-inference} identifies and rewrites leaked benchmark samples while preserving their complexity.

\subsection{LLM as Judgment Agent}

Recently, LLMs have become increasingly used as examiners given their capabilities of analyzing large amounts of data and providing unbiased assessments \cite{bai2023benchmarking, fernandes2023devilerrors}. This growing trend has gained interest for two reasons: (1) Enhanced generation of training/testing data \cite{li2024treeeval, liu2024augmentingmathwordproblems} (2) Accurate evaluation and comparison of LLM outputs such as in PandaLM \cite{wang2024pandalm} and DyVal \cite{zhu2024dynamicevaluation}. Additionally, as LLMs have been able to perform remarkbly well on unseen tasks, they offer a faster, equally accurate alternative to human evaluation,  \cite{chiang2023largelanguagemodelsalternative}.

\section{Methods: \tool}

\subsection{Design Overview}

There are two key challenges in designing a dynamic evaluation schema for benchmarking code LLMs. (1) \textit{Generating Semantically Diverse yet Complexity-Controlled Problems}: There is currently no systematic method for generating programming problems that maintain a consistent complexity level while ensuring semantic diversity. Existing approaches often rely on manual effort, either through predefined rules or domain experts, making them difficult to scale efficiently and incapable of precisely controlling problem complexity.  
(2) \textit{Ensuring Comprehensive Benchmarking}: To effectively evaluate code LLMs, the generated programming problems must include fine-grained test cases and canonical solutions to rigorously assess correctness.  


We draw inspiration from metamorphic testing to generate programming problems using LLMs as agents. Metamorphic testing, widely used in software engineering, defines relationships to address the automatic oracle problem.  
In our approach, a programming problem prompt consists of two components: \textit{complexity-related algorithmic abstraction} and \textit{complexity-unrelated context description}. Our key metamorphic relationship states that modifying the \textit{complexity-unrelated context description} preserves both the problem’s canonical solutions and complexity, enabling controlled problem generation.  
Additionally, since LLMs are trained on a vast diverse corpus, we can utilize them as agents to suggest relevant and meaningful \textit{complexity-unrelated context descriptions}, further enhancing problem diversity.

\begin{figure*}
    \centering
    \includegraphics[width=0.88\textwidth]{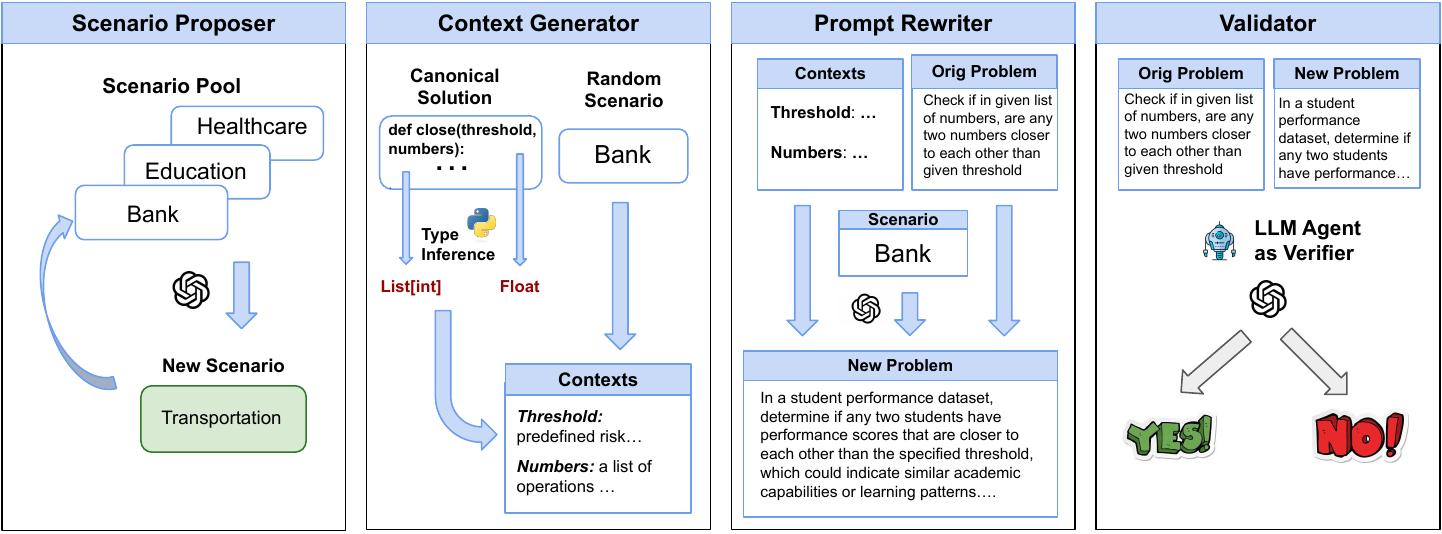}

    \caption{Design overview of \tool}
    \label{fig:overview}
\end{figure*}

The design overview of \tool is shown in \figref{fig:overview}. Given a seed programming problem from existing benchmarks, \tool generates a semantically different yet complexity-equivalent problem using a metamorphic relationship. \tool comprises of four agents: (1) Scenario Proposer, (2) Context Generator, (3) Prompt Rewriter, and (4) Validator. The Scenario Proposer suggests real-world domains (e.g., banking, healthcare, education) from which \tool randomly selects one. The Context Generator then analyzes input types in the canonical solution and assigns a relevant context for each input variable based on the selected scenario. The Prompt Rewriter reformulates the problem to align with the input variable contexts and chosen scenario. Finally, the Validator ensures the new problem remains consistent with the original. If inconsistencies are detected, \tool will repeat the aforementioned process until a valid variant is produced.

\subsection{Detailed Design}
\fakeparagraph{Scenario Proposer Agent} The Scenario Proposer enhances diversity and minimizes repetition in generated programming problems, reducing potential data contamination. It first selects scenarios from a predefined pool (e.g., banking, healthcare, education, transportation, social networking) and uses them as examples to prompt an LLM for new scenario suggestions. The newly generated scenarios are then added to the pool. By iteratively updating the pool and querying the LLM with varied examples, \tool continuously expands the scenario diversity until the scenario pool reaches a pre-defined size, ensuring the generated scenarios remain diverse and practical. The prompt used for querying the LLM and the suggested scenario examples are listed in \appref{appendix:prompt_templates}.

\begin{algorithm}[tbp!]
\caption{Type Inference Algorithm.  $\quad$ \texttt{Abstract} ($\cdot$)} 
\label{alg:alg}

\begin{flushleft}
 {\bf Input:} Value list $\mathcal{V}$. \\
 {\bf Output:} Set of data types  $\vec{\tau}$. \\
\end{flushleft}

\begin{algorithmic}[1]

\STATE $\vec{\tau}$ = \{$\;$\}  $\quad\quad\quad\quad\quad\quad\quad\quad\quad\quad\quad\quad$ // Initialization.

\FOR {each v $\text{in} \; \mathcal{V}$}
    \STATE $\tau$ = \texttt{Type}(v)
    \IF {$\tau$ $\in$ Basic Types}
        \STATE \label{alg:basic} $\vec{\tau} = \vec{\tau}.add(\texttt{Type}(v)$)   
    \ELSE
        \STATE $\tau^* = \texttt{Abstract}(\texttt{ToList}(v))$ $\quad\quad$ 
        \STATE $\vec{\tau}.add(\tau[\tau^*])$  $\quad\quad\quad\quad\quad\quad\quad$ // Composite type.
        
    \ENDIF
\ENDFOR
\STATE return $\vec{\tau}$
\end{algorithmic}
\end{algorithm}

\fakeparagraph{Context Generation Agent}
After proposing a set of scenarios, the context generation agent randomly selects one from the pool and assigns context information to each input variable of the programming problem based on the chosen scenario.  

In languages like Python, input types are not explicitly defined. To address this, the agent uses abstraction for type inference. It analyzes \texttt{ASSERT} statements in test cases, collects concrete input values from the canonical solution, and abstracts the input type based on these values. Our type inference algorithm, shown in \algref{alg:alg}, works as follows: for each concrete value, it first checks if the type is a basic type (\eg \texttt{int}, \texttt{float}). If so, it updates the type set. Otherwise the value is a composite type so it recursively iterates over all the elements and updates the type set with types like \texttt{List[int]} or \texttt{Tuple[int | string]}. Notice that while our abstract-based type inference may not capture all return value types, it is sound and guarantees that the collected types will always appear in the canonical solution.

After collecting the input data types, the agent prompts the LLM with the scenario and input type information, asking it to assign meaningful context to each input variable based on the given scenario. See \appref{appendix:prompt_templates} for prompt templates of our context generation.

\fakeparagraph{Prompt Rewriting Agent} 
With the scenario and context information for each input variable, the prompt rewriting agent then rewrites the seed programming problem prompt to be tailored to the scenario with meaningful context. Note that we did not ask the LLM to generate the new prompt from scratch. Instead, we provided the detailed scenario and asked it to perform a rewriting task, which is simpler than a generation task. With this approach, leveraging detailed context and a more straightforward task, our agent can generate semantically diverse programming problem prompts. See \appref{appendix:prompt_templates} for prompt templates of our prompt rewriting.

\fakeparagraph{Validation Agent}
Although we provide the LLM with detailed scenario and context information for rewriting, there are cases where the rewriting agent unintentionally alters the consistency. To address this, we design a validation agent to assess whether the generated question maintains the integrity of the original intent and informational content. The validation prompt is designed from two angles: (1) it directs the LLM to compare the seed programming problem prompt with the rephrased prompt, ensuring the preservation of the core concept and factual accuracy, and (2) it asks the LLM to check whether the seed canonical solutions align with the generated programming problem prompt. Specifically, we design two comparison prompts to query the LLM and retain only those rewritten prompts for which both comparison responses are ``YES''.

\begin{figure}
    \centering
    \includegraphics[width=0.48\textwidth]{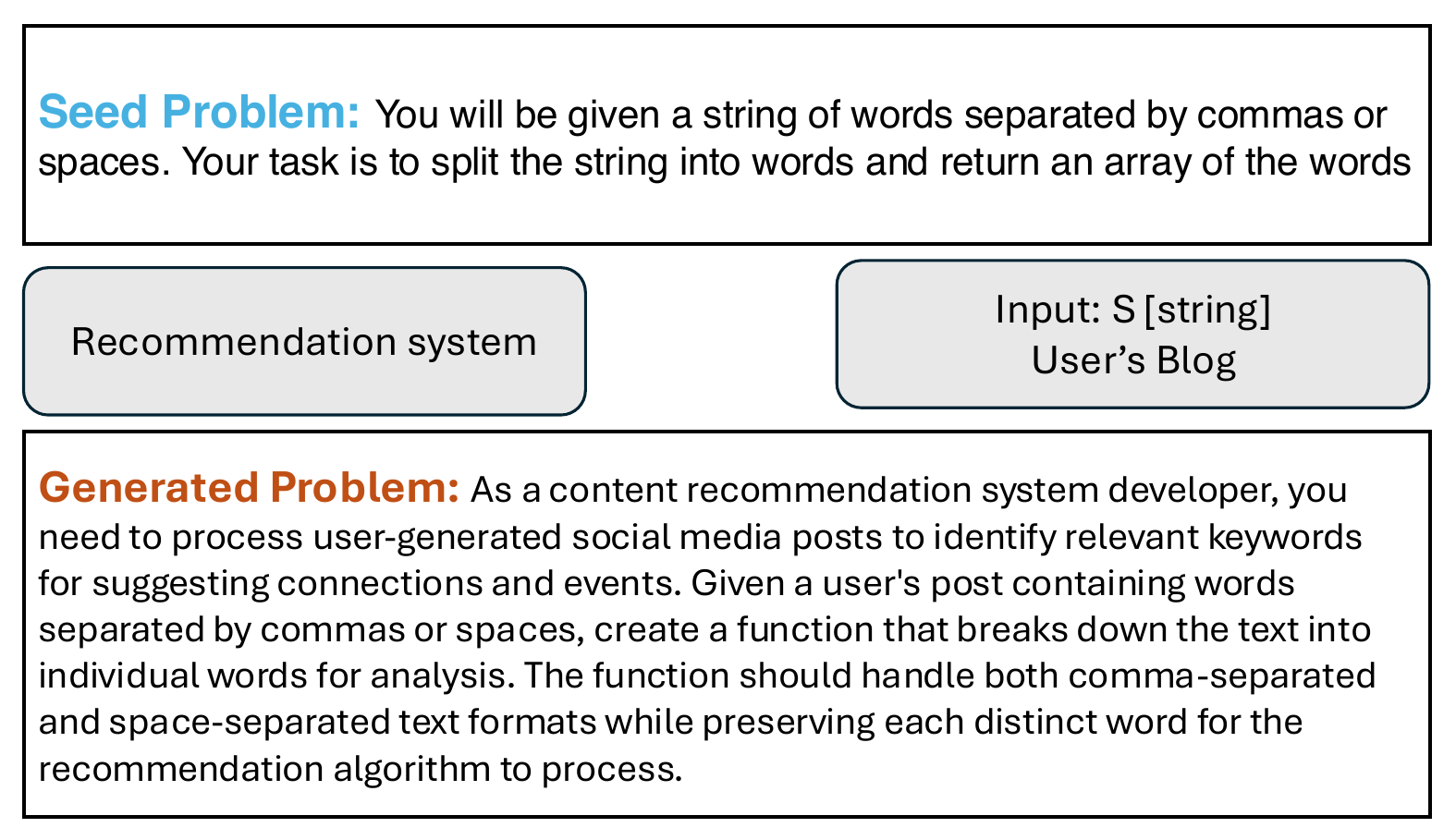}
    
    \caption{A generated example from \tool}
    \label{fig:example}
\end{figure}

To ensure the consistency of the generated programming problems, we also include a human verification step. The details of our validation prompt and the human verification process are presented in \appref{appendix:prompt_templates} and \appref{appendix:human_verification}.

\figref{fig:example} illustrates an example of programming problems that are semantically diverse yet complexity-equivalent, generated under the scenario of a recommendation system with the context of a user's blog. From this example, we observe that our step-by-step guided approach significantly enhances the semantic diversity of the generated problems, while also reducing the risk of data contamination. This is achieved by leveraging the vast combination space of scenarios and contexts.

\subsection{Theoretical Collision Analysis}

\tool generates programming problems dynamically with randomness, reducing the risk of potential data contamination. To analyze this, we conduct a collision analysis. The randomness in \tool arises from both the scenario proposal and context generation phases. We assume the scenario proposer generates $||\mathcal{S}||$ scenarios, and for each scenario, the context generation produces $||\mathcal{C}||$ contexts, while ignoring randomness in the rewriting phase. 
We also assume that the random sampling process follows a uniform distribution.
Based on this, we present the following theorem.

\begin{theorem}
\label{theorem:1}
After running \tool $M + 1$ times on the same seed problem, then the probability that the $M$ samples after the first are all different from the first sampled item satisfies: $P \geq 1 - \exp\left(-\frac{M}{||\mathcal{S}|| \times ||\mathcal{C}|| - 1}\right)$.
\end{theorem}

\begin{theorem}
\label{theorem:2}
After running \tool $M$ times on the same seed problem, If $M << ||\mathcal{S}|| \times ||\mathcal{C}||$, the probability of at least one collision (i.e., two or more generated problems being the same) after $M$ generations satisfies the following bound: $P \leq 1 - \exp\left(-\frac{M^2 - M}{2||\mathcal{S}|| \times ||\mathcal{C}||}\right)$.
\end{theorem}

\begin{theorem}
\label{theorem:3}
Consider the seed dataset of size $\mathcal{D}$, After running \tool $M + 1$ times on this dataset, If $M << ||\mathcal{S}|| \times ||\mathcal{C}||$, then the probability that the $M$ generated datasets after the first one are not the same as the first generated dataset satisfies: $
1 - e^{-\frac{M}{(||\mathcal{S}|| \times ||\mathcal{C}|| )^\mathcal{D} -1 }} \leq P $
\end{theorem}

The proof could be found in \appref{appendix:proof}.

\begin{figure*}[thbp!]
    \centering
    \includegraphics[width=0.96\textwidth]{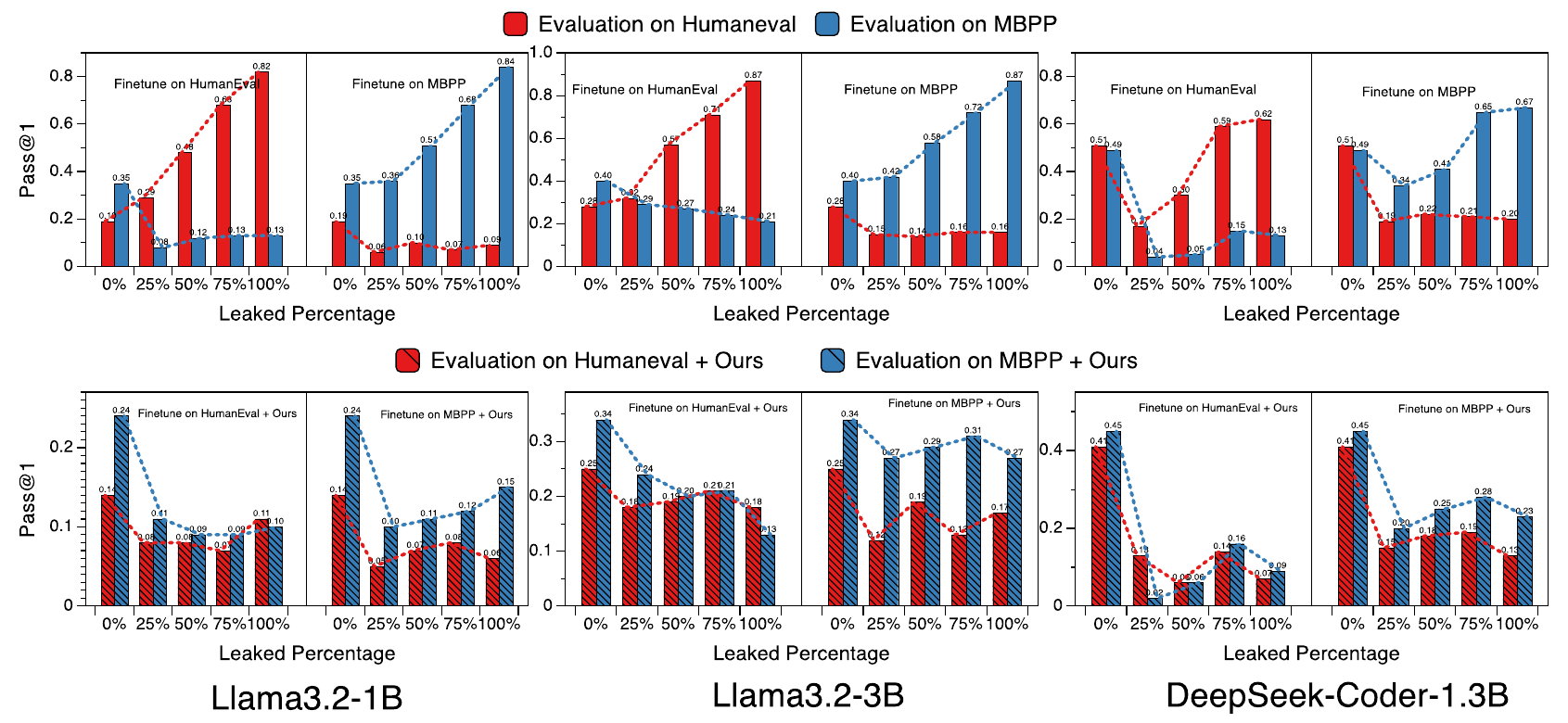}

    \caption{Results of benchmarking on contaminated models}
    \label{fig:contamination}
\end{figure*}

\section{Evaluation}

\subsection{Experimental Setup}

\fakeparagraph{Seed Dataset} We conduct our evaluation using two datasets: \textit{HumanEval} \cite{humaneval} and \textit{MBPP-Sanitized} \cite{mbpp}. Both datasets are widely utilized in existing research and serve as standard benchmarks for evaluating code generation models.
More details about the dataset could be found in \appref{appendix:dataset}.

\fakeparagraph{Implementation Details} 
We use \textsc{Claude-3.5-Sonnet} as our foundation model to generate the benchmarking dataset. Specifically, we create 50 scenarios, and for each scenario, we randomly generate 50 contexts. During dataset generation, we set the LLM temperature to 0.8, while in our validation agent, we use a temperature of 0. For each code LLM under benchmarking, we employ \texttt{vLLM} to launch the model. For closed-source code LLMs, we query the commercial API for evaluation.

\subsection{Benchmarking Contaminated Model} 
\label{sec:contaminated}
\fakeparagraph{Models} We conduct our study with three public-available Code LLMs: \textsc{Llama-3.2-1B}, \textsc{Llama-3.2-3B}, and \textsc{DeepSeek-Coder-1.3b}. 

\fakeparagraph{Model Contamination Process} 
For each model, we simulate data contamination by intentionally leaking a portion of the benchmarking dataset during fine-tuning. 
We experiment with leaked data percentages of 0\%, 25\%, 50\%, 75\%, and 100\%, producing four distinct contaminated models. Each polluted model is then evaluated on the benchmarking dataset using the \texttt{Pass@1} metric. The formal definition of \texttt{Pass@1} is shown in \eqref{eq:pass@k}, where $n$ is the number of the generated solution candidate, and $c$ is the number of the correct solutions that can pass all test cases. 

\begin{equation}
\label{eq:pass@k}
    \texttt{Pass@K} = \mathbb{E}_{\text{Problems}} \left[ 1 - \frac{\binom{n-c}{k}}{\binom{n}{k}} \right]
\end{equation}


\fakeparagraph{Main Results}
The study results are presented in \figref{fig:contamination}, where there are two rows and three columns. Each column represents evaluation on a different LLM while the rows show static (first) vs dynamic (second) benchmarking. In each column, the left section displays the results for the model fine-tuned on the \textit{HumanEval} dataset, while the right section shows the results for the model fine-tuned on the \textit{MBPP} dataset. The red bars represent the performance of the fine-tuned model benchmarked on the \textit{HumanEval} dataset, and the blue bars represent its performance benchmarked on the \textit{MBPP} dataset.

From the results, we make the following observations:  (1) Data contamination creates a false sense of code reasoning capability under static benchmarks. When the benchmarking dataset is leaked and used for fine-tuning, the model achieves a higher \texttt{Pass@1} score on the corresponding benchmark. However, this improvement does not accurately reflect the model’s true reasoning ability, as its performance declines on other benchmarks that were not included in fine-tuning.  (2) Our dynamic benchmarking mitigates the impact of data contamination. Different from static benchmarks, our approach prevents contaminated models from achieving artificially high \texttt{Pass@1} scores after fine-tuning. This is due to the randomness in our method, which ensures minimal or no overlap between different runs, reducing the risk of direct data leakage.  (3) Our dynamic benchmarking dataset provides results comparable to manually curated, non-contaminated datasets. In static benchmarking, as the percentage of leaked data increases, the model’s \texttt{Pass@1} score on the contaminated benchmark steadily improves. However, its performance on other benchmarks remains relatively stable, showing little variation across different contamination levels. Interestingly, this stability also applies to our method. If the base model is not contaminated on the selected seed dataset, this suggests that our approach provides competitive benchmarking results similar to those of human-curated datasets.  (4) A notable anomaly is observed in \textsc{DeepSeek-Coder}. When only 25\% of the benchmarking dataset is used for fine-tuning, the model’s \texttt{Pass@1} score drops below that of the original, unmodified model. We hypothesize that the model may already be overfitted to the contaminated dataset, and further fine-tuning with limited data could destabilize this overfitting without providing enough new information to help the model adapt.


\begin{figure*}[thbp!]
    \centering
    \includegraphics[width=0.88\linewidth]{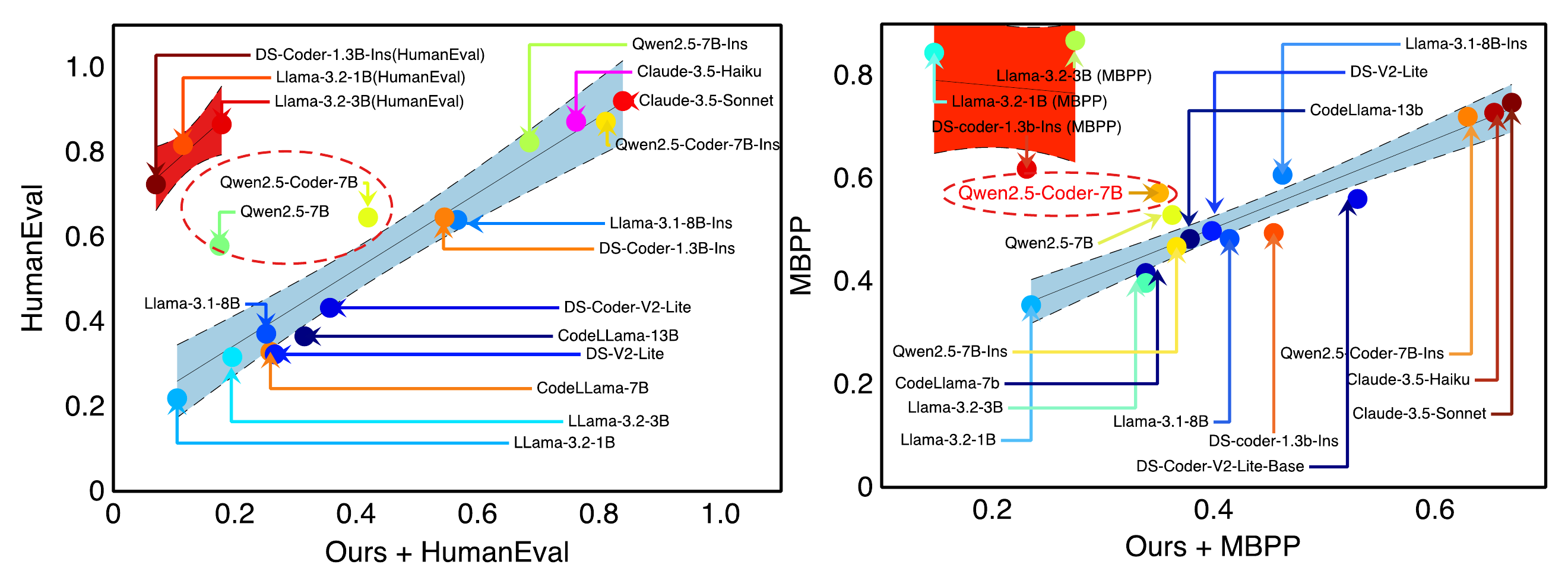}
    \caption{The in-the-wild benchmarking results}
    \label{fig:wild}
\end{figure*}

\subsection{Benchmarking In-the-Wild Model}
\label{sec:wild}

We then apply \tool to benchmark more in-the-wild code LLMs, besides the models used in \secref{sec:contaminated}. We consider the following code LLMs: \textsc{Llama-3.1-8B}, 
\textsc{CodeLlama-7b}, \textsc{CodeLlama-13b}, \textsc{DeepSeek-V2-Lite}, \textsc{DeepSeek-Coder-V2-Lite-Base}, \textsc{Llama-3.1-8B-Instruct},  
\textsc{Qwen2.5-Coder-7B}, \textsc{Qwen2.5-7B-Instruct}, \textsc{Qwen2.5-7B}, \textsc{Claude-3.5-haiku}, \textsc{Claude-3.5-sonnet},\textsc{Qwen2.5-Coder-7B-Instruct} .

The results are presented in \figref{fig:wild}, with the left figure showing the results on HumanEval and the right showing the results on MBPP. In each figure, the x-axis represents the \texttt{Pass@1} scores on our generated dataset, and the y-axis represents the \texttt{Pass@1} scores on the seed dataset. The blue region corresponds to the regression area of the in-the-wild model, the red region represents the regression area of the overfitted model on this dataset, and the orange area indicates the overfitted model on the other dataset.  

From these results, we observe that for both seed datasets, the in-the-wild model’s \texttt{Pass@1} scores maintain a linear relationship, while the overfitted model appears as an outlier. A notable finding from our in-the-wild evaluation is that the model \textsc{Qwen2.5-Coder-7B} consistently falls outside the 95\% confidence interval of the regression area, suggesting it may be contaminated on both datasets.

\subsection{Problem Diversity}
\label{sec:diversity}

To evaluate the diversity of the generated programming problems, we conduct two experiments: one for external diversity and one for internal diversity. External diversity quantifies the dissimilarity between the generated and seed problems, while internal diversity measures the diversity within each problem-generation method across trials. We use two metrics: \textit{BLEU-4} to measure syntactical diversity and \textit{cosine similarity} of the prompt’s semantic embedding to measure semantic diversity. For semantic embedding, we use the GPT-2 model to obtain the embedding of each natural language prompt.
Moreover, we also consider \texttt{PPM} \cite{chen2024ppm} and a series of robustness-based mutations~\cite{recode}, such as token replacement, insert blank lines, as our comparison baseline.

The diversity results are shown in \tabref{tab:diversity}, where the first four columns represent internal diversity and the last four columns represent external diversity. From the results, we observe that \tool generates diverse programming problems both syntactically and semantically. Additionally, we find that all baseline methods exhibit high BLEU-4 and semantic similarity scores, as they rely on rule-based approaches to mutate the programming problems, which do not introduce significant diversity. In contrast, \tool leverages an LLM agent to suggest different scenarios and contexts, significantly increasing diversity.

\begin{table*}[htbp!]
  \centering
  \caption{Diversity results}
    \resizebox{0.8\textwidth}{!}{
    \begin{NiceTabular}{l|cc|cc|cc|cc}
    \CodeBefore
        \rowcolors{3}{}{gray!12}
        \Body
    \toprule
    \toprule
    \multicolumn{1}{c}{\multirow{3}[2]{*}{\textbf{Methods}}} & \multicolumn{4}{c}{\textbf{Internal Diversity}} & \multicolumn{4}{c}{\textbf{External Diversity}} \\
          & \multicolumn{2}{c}{\textbf{HumanEval}} & \multicolumn{2}{c}{\textbf{MBPP}} & \multicolumn{2}{c}{\textbf{HumanEval}} & \multicolumn{2}{c}{\textbf{MBPP}} \\
          & \boldmath{}\textbf{BLEU-4 $\downarrow$}\unboldmath{} & \boldmath{}\textbf{SemSim $\downarrow$}\unboldmath{} & \boldmath{}\textbf{BLEU-4 $\downarrow$}\unboldmath{} & \boldmath{}\textbf{SemSim $\downarrow$}\unboldmath{} & \boldmath{}\textbf{BLEU-4 $\downarrow$}\unboldmath{} & \boldmath{}\textbf{SemSim $\downarrow$}\unboldmath{} & \boldmath{}\textbf{BLEU-4 $\downarrow$}\unboldmath{} & \boldmath{}\textbf{SemSim $\downarrow$}\unboldmath{} \\
    \midrule
    \textbf{Base} & 1.00  & 1.00  & 1.00  & 1.00  & 1.00  & 1.00  & 1.00  & 1.00  \\
    \midrule
    \textbf{Token Mutation} & 0.72  & 0.95  & 0.66  & 0.92  & 0.82  & 0.96  & 0.76  & 0.95  \\
    \textbf{Char Mutation} & 0.81  & 0.97  & 0.78  & 0.94  & 0.84  & 0.97  & 0.78  & 0.92  \\
    \textbf{Func Mutation} & 1.00  & 1.00  & 1.00  & 1.00  & 0.98  & 1.00  & 0.98  & 1.00  \\
    \textbf{Insert Line} & 1.00  & 1.00  & 1.00  & 1.00  & 1.00  & 1.00  & 1.00  & 1.00  \\
    \textbf{CommSyntax} & 1.00  & 1.00  & 1.00  & 1.00  & 0.81  & 0.98  & 0.73  & 0.99  \\
    \textbf{PPM} & 0.97  & 0.96  & 0.96  & 0.94  & 0.69  & 0.89  & 0.57  & 0.84  \\
    \midrule
    \textbf{Ours} & \textbf{0.27 } & \textbf{0.74 } & \textbf{0.18 } & \textbf{0.73 } & \textbf{0.17 } & \textbf{0.59 } & \textbf{0.02 } & \textbf{0.59 } \\
    \bottomrule
    \bottomrule
    \end{NiceTabular}%
    }
  \label{tab:diversity}%
\end{table*}%

\begin{figure}[hbp!]
    \centering
    \includegraphics[width=0.46\textwidth]{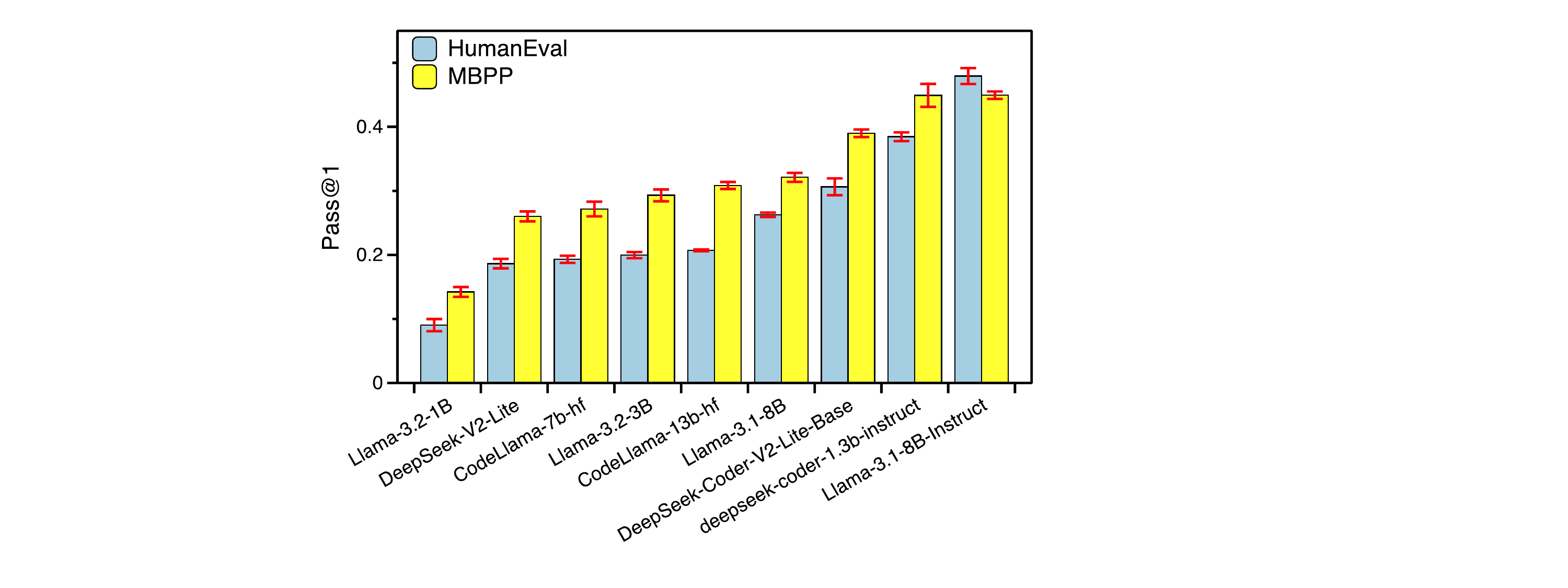}
    \caption{Stability results}
    \label{fig:stable}
\end{figure}

\subsection{Benchmarking Stability}
\label{sec:stable}
Note that \tool generates a unique benchmarking dataset each time. To assess its stability, we evaluate whether \tool can produce consistent benchmarking results despite this randomness. Specifically, we run \tool 10 times and measure the \texttt{Pass@1} scores across these 10 generated benchmark datasets.

The mean and standard deviation of the \texttt{Pass@1} scores are presented in \figref{fig:stable}. The results show that the variance in benchmarking scores is minimal compared to the mean values, indicating that \tool provides stable benchmarking results across different random trials.

\subsection{Impact of Foundation LLM}

In this section, we evaluate the feasibility of using less advanced LLMs to reduce dataset generation costs. Specifically, we replace our foundation model, \textsc{Claude-3.5-Sonnet}, with \textsc{Claude-3.5-Haiku}. We manually sample and assess generated problems from each model, checking their consistency rate. Our observations show that the consistency rate drops from 95\% to 83\%, highlighting the need for robust and capable LLMs to serve effectively as foundation models.




\section{Dynamic Evaluation Metrics}

Leveraging the dynamic nature of our method, we propose a new metric, \metric, to address the limitations of the current gold standard, \texttt{Pass@K}. Unlike \texttt{Pass@K}, which generates \( n \) candidate solutions for a fixed problem prompt and evaluates the correctness, our approach creates \( n \) semantic prompt variants of a seed problem. These prompt variants preserve the complexity of the original problem by modifying only the description while maintaining the same underlying algorithmic abstraction.  
Furthermore, \( n \) prompt variants expand the input space beyond that of \texttt{Pass@K}, making it more challenging to achieve full coverage. As a result, \metric provides a more rigorous assessment of code LLMs' reasoning abilities, particularly under potential data contamination. Compared to \texttt{Pass@K}, which evaluates solutions within a fixed problem context, \metric introduces contextual variations during benchmarking. This allows it to better distinguish whether a model is merely memorizing the problem context or genuinely reasoning to solve it.

\begin{table}[htbp]
  \centering
  \caption{Comparison of \texttt{Pass@K} and \texttt{DyPass@K} on contaminated Models}
  \resizebox{0.48\textwidth}{!}{
    \begin{NiceTabular}{lcccccc}
    \CodeBefore
        \rowcolors{3}{}{gray!12}
        \Body
    \toprule
    \toprule   
    \multirow{2}[2]{*}{Model} & \multicolumn{3}{c}{\textcolor[rgb]{ .275,  .471,  .525}{Pass@K}} & \multicolumn{3}{c}{\textcolor[rgb]{ .275,  .471,  .525}{DyPass@K}} \\
          & k=3   & k=5   & k=10  & k=3   & k=5   & k=10 \\
    \midrule
    Llama-3.2-1B & 0.22  & 0.27  & 0.34  & 0.17  & 0.21  & 0.26 \\
    Llama-3.2-1B (C) & 0.82  & 0.83  & 0.85  & 0.13  & 0.15  & 0.17 \\
    \midrule
    Llama-3.2-3B & 0.35  & 0.40  & 0.48  & 0.31  & 0.36  & 0.43 \\
    Llama-3.2-3B (C) & 0.88  & 0.88  & 0.89  & 0.24  & 0.27  & 0.29 \\
    \bottomrule
    \bottomrule
    \end{NiceTabular}%
    }
  \label{tab:div_pass_con}%
\end{table}%

\begin{table}[htbp]
  \centering
  \caption{Comparison of \texttt{Pass@K} and \texttt{DyPass@K} on in-the-wild models}
  \resizebox{0.48\textwidth}{!}{
    \begin{NiceTabular}{lcccccc}
    \CodeBefore
        \rowcolors{3}{}{gray!12}
        \Body
    \toprule
    \toprule
    \multirow{2}[2]{*}{Model} & \multicolumn{3}{c}{\textcolor[rgb]{ .275,  .471,  .525}{Pass@K}} & \multicolumn{3}{c}{\textcolor[rgb]{ .275,  .471,  .525}{DyPass@K}} \\
          & k=3   & k=5   & k=10  & k=3   & k=5   & k=10 \\
    \midrule
    CodeLlama-7b-hf & 0.39  & 0.46  & 0.56  & 0.34  & 0.40  & 0.49 \\
    CodeLlama-13b-hf & 0.48  & 0.57  & 0.68  & 0.37  & 0.45  & 0.53 \\
    \midrule
    Llama-3.2-1B & 0.22  & 0.27  & 0.34  & 0.17  & 0.21  & 0.26 \\
    Llama-3.2-3B & 0.35  & 0.40  & 0.48  & 0.31  & 0.36  & 0.43 \\
    Llama-3.1-8B & 0.48  & 0.56  & 0.65  & 0.39  & 0.45  & 0.53 \\
    Llama-3.1-8B-Instruct & 0.72  & 0.77  & 0.83  & 0.64  & 0.69  & 0.75 \\
    \bottomrule
    \bottomrule
    \end{NiceTabular}%
    }
  \label{tab:div_pass}%
\end{table}%

To demonstrate the advantages of \metric, we compare it against \texttt{Pass@K} on both contaminated and in-the-wild models, with \( K = 3, 5, 10 \) for evaluation. The results are presented in \tabref{tab:div_pass_con} and \tabref{tab:div_pass}. From the results in \tabref{tab:div_pass_con}, we observe that when the model is trained on leaked data, the static metric \texttt{Pass@K} fails to accurately reflect the model's reasoning capabilities, with all \texttt{Pass@K} scores rising to very high levels (e.g., from 0.82 to 0.89). In contrast, our dynamic metric \metric@K shows a slight decrease rather than a significant increase, highlighting the sensitivity of \metric to data contamination.   When comparing \texttt{Pass@K} and \metric@K on models that were not specifically trained on the leaked dataset, both metrics show consistency in benchmarking code LLMs. Based on these observations, we conclude that our dynamic metric, \metric, effectively reflects the reasoning capabilities of code LLMs, even under data contamination. Moreover, \metric@K aligns with static benchmarking metrics when there is no data contamination.




\section{Conclusion}

In this paper, we introduce \texttt{DyCodeEval}, a new benchmarking suite that dynamically generates semantically equivalent diverse problems as a way to combat data contamination. We break this generation up into four distinct steps to systematically develop a new programming problem with the same algorithmic complexity but different context. Our experimental results show that while \texttt{Pass@k} with current benchmarks have caused inflated model scores, \texttt{DyCodeEval}-generated questions with \texttt{DivPass} has proven to perform as a reliable evaluation tool. We believe that these results show a promising path forward.

Our proposed work has several limitations: (1) Although LLMs provide a fully automated way to generate diverse programming problems for benchmarking, their computational cost is a significant concern. We found that a very large LLM is required to generate programming problems with a high consistency rate. Therefore, a future improvement could focus on enhancing the efficiency of the problem generation phase. (2) While generating questions using DyCodeEval, we observed instances where excessive information was provided, potentially confusing the reader. This highlights the opportunity for improving prompt generation through further experimentation.


\section*{Acknowledgements}
This work was supported in part by CCF 2313055, CCF 2107405, CAREER 2025082, and FAI: 2040961.  Any opinions, findings, conclusions, or recommendations expressed herein are those of the authors.


\section*{Impact Statement}




Assessing the overall capabilities of large language models (LLMs) is essential for ensuring their reliable and safe deployment in society. However, data contamination can inflate evaluation accuracy, obscuring a model’s true performance. To address this, we propose a new benchmarking method, \tool, which enables more accurate measurement of LLM capabilities and provides deeper insights into their behavior.




\bibliography{example_paper}
\bibliographystyle{icml2025}

\newpage
\appendix
\onecolumn

\section{Proof of Theorem}
\label{appendix:proof}

\subsection{Proof of Theorem \ref{theorem:1}}

The total number of possible distinct outcomes is $||\mathcal{S}|| \times ||\mathcal{C}||$, the size of the random space, let $N = ||\mathcal{S}|| \times ||\mathcal{C}||$
Since each of the \( M \) samples must \textbf{not} match \( X_1 \), and they are drawn independently, the exact probability is:

\[
P(X_2 \neq X_1, \dots, X_{M+1} \neq X_1) = \left( \frac{N-1}{N} \right)^M.
\]

We use the standard inequality for the logarithm:

\[
\ln(1 - x) \geq -\frac{x}{1-x}, \quad \text{for } 0 < x < 1.
\]

Applying this to \( \frac{1}{N} \), we get:

\[
\ln \left( \frac{N-1}{N} \right) = \ln \left( 1 - \frac{1}{N} \right) \geq -\frac{1/N}{1 - 1/N} = -\frac{1}{N-1}.
\]

Exponentiating both sides:

\[
\frac{N-1}{N} \geq e^{-\frac{1}{N-1}}.
\]

Raising both sides to the power \( M \):

\[
\left( \frac{N-1}{N} \right)^M \geq e^{-\frac{M}{N-1}}.
\]

\subsection{Proof of Theorem \ref{theorem:2}}

Each sampled item is drawn independently and uniformly from the space of size \( N \). We analyze the probability that all \( M \) sampled items are distinct.

The first sample can be any of the \( N \) items, the second sample must avoid the first one, so there are \( N-1 \) choices. Continuing this way, the probability that all \( M \) items are distinct is:

\[
P(\text{no collisions}) = \frac{N}{N} \times \frac{N-1}{N} \times \frac{N-2}{N} \times \cdots \times \frac{N - (M-1)}{N}.
\]

Rewriting in factorial form,

\[
P(\text{no collisions}) = \frac{N!}{N^M (N-M)!}.
\]

According to our assumption $M << ||\mathcal{S}|| \times ||\mathcal{C}||$, Using the Stirling’s approximation, then we have 

\[
\frac{N!}{(N-M)!} \ge N^M \exp\left(-\frac{M(M-1)}{2N}\right),
\]

we get

\[
P(\text{no collisions}) \ge  \exp\left(-\frac{M(M-1)}{2N}\right).
\]

The probability of at least one collision is the complement:

\[
P(\text{at least one collision}) = 1 - P(\text{no collisions}).
\]

Using the bound we derived,

\[
P(\text{at least one collision}) \leq 1 - \exp\left(-\frac{M^2 - M}{2N}\right) = 1 - \exp\left(-\frac{M^2 - M}{2 ||\mathcal{S}|| \times ||\mathcal{C}||}\right)
\]

\subsection{Proof of Theorem \ref{theorem:3}}

Each sample can be represented as a \( D \)-tuple of balls \((b_1, b_2, ..., b_D)\), where each \( b_i \) is one of the \( N \) balls from bag \( i \). The total number of possible sample sets is:

\[
T = N^D
\]

Since each draw is independent, each sample set is chosen uniformly from \( T \), meaning the probability of selecting any specific tuple is:

\[
\frac{1}{N^D}
\]

Let \( X_1 \) be the initial sample (first draw). For each subsequent draw \( X_i \) (where \( i = 2, \dots, M+1 \)), the probability that \( X_i = X_1 \) (i.e., an exact match) is:

\[
P(X_i = X_1) = \frac{1}{N^D}
\]

Then Theorem \ref{theorem:3} could be proved through Theorem \ref{theorem:1}.

\section{Dataset Description.}
\label{appendix:dataset}
The \textit{HumanEval} dataset, developed by OpenAI, is an open-source benchmark for evaluating the code generation capabilities of pre-trained code language models (LLMs). It comprises 164 Python programming problems, each consisting of a prompt, a canonical solution, and corresponding test inputs. Each prompt includes a natural language problem description, a function definition, and input/output examples.

The \textit{MBPP-Sanitized} dataset, proposed by Google, features 427 Python programming problems collected through crowdsourcing. Unlike \textit{HumanEval}, it is a zero-shot dataset, meaning its prompts do not include input/output demonstrations. To enhance its utility in experiments, we refined the prompt format by adding function headers and converting natural language instructions into function docstrings.
\section{Prompt Templates \& Scenario Examples}
\label{appendix:prompt_templates}

In the following, we show the scenario examples and prompt templates used during the four steps of \texttt{DyCodeEval} process.




\subsection{Template for Scenario Proposer Agent}
\begin{tcolorbox}[
    colback=gray!10!white,        
    colframe=black,               
    title=Prompt for Scenario Proposer Agent,
    fonttitle=\bfseries,
    coltitle=white,              
    colbacktitle=black,          
    boxrule=0.8pt,               
    arc=1mm,                     
    enhanced                    
]
\ttfamily 
Suggest real-world scenarios that provide meaningful context in the following areas: \textbf{\{$S_1$\}, \{$S_2$\}, \{$S_3$\}, \{$S_4$\}, \{$S_5$\}}, and any other practical fields. Each scenario should be general but applicable, providing useful insight for potential applications. \\

For clarity, return each scenario on a separate line without additional explanation. Use the example below for reference. \\

Please put your suggested Real-world Scenarios in <scenario></scenario> tags.
 \\

\# Scenario Examples: \\
$<$example$>$ \\
\textbf{\{EXAMPLE\}} \\
$</$example$>$\\
\newline
\end{tcolorbox}

\subsection{Example for Scenario Proposer Agent}
\begin{tcolorbox}[
    colback=orange!10!white,        
    colframe=orange,               
    title=Example for Scenario Proposer Agent,
    fonttitle=\bfseries,
    coltitle=white,              
    colbacktitle=orange,          
    boxrule=0.8pt,               
    arc=1mm,                     
    enhanced                    
]
\ttfamily 
Suggest real-world scenarios that provide meaningful context in the following areas: \textcolor{pyblue}{transportation, education, healthcare, banking, social networking}, and any other practical fields. Each scenario should be general but applicable, providing useful insight for potential applications. \\

For clarity, return each scenario on a separate line without additional explanation. Use the example below for reference. \\

Please put your suggested Real-world Scenarios in <scenario></scenario> tags.
 \\

\# Scenario Examples: \\
$<$example$>$ \\
\textcolor{pyred}{Banking - Fraud Detection.} \\
$</$example$>$\\
\newline

\end{tcolorbox}

\subsection{Prompt for Context Generator Agent}
\begin{tcolorbox}[
    colback=gray!10!white,        
    colframe=black,               
    title=Prompt for Content Generator Agent,
    fonttitle=\bfseries,
    coltitle=white,              
    colbacktitle=black,          
    boxrule=0.8pt,               
    arc=1mm,                     
    enhanced,
    fontupper=\small
]
\ttfamily 
Given the natural language problem description, input types, and a real-world scenario. For each variable in the problem, provide a meaningful context tailored to the given scenario. The context should explain how each variable is involved in or relates to the scenario, ensuring practical relevance. \\ 

Please ensure for to put your meaningful context in <context></context> tags. \\

\# Problem Description: \\
$<$problem\_description$>$  \\
\textbf{\{PROBLEM DESCRIPTION\}} \\
$<$/problem\_description$>$ \\

\# Input Types: \\ 
$<$input\_types$>$  \\
\textbf{\{INPUT VARIABLE TYPES\}}   \\ 
$<$/input\_types$>$  \\

\# Real-world Scenario: \\
$<$scenario$>$          \\
\textbf{\{SCENARIOS\}}   \\
$<$/scenario$>$          \\

\# Instructions:          \\
- For each variable in the input types, generate only one context that highlights its role or significance within the problem and scenario. \\
- The context should help to clarify the variable’s meaning and importance, ensuring that it fits into the given real-world scenario. \\ 
- Provide only the contexts for the variables (no additional reasoning steps).  \\

\# Example:  \\
\#\# Problem Description Example:  \\ 
$<$problem\_description$>$         \\
Determine if the average temperature in a city exceeds a certain threshold during a week.  \\
$<$/problem\_description$>$         \\

\#\# Input Types:         \\
$<$input\_types$>$  \\
temperatures: list of float  \\
threshold: float   \\
$<$/input\_types$>$  \\

\#\# Scenario:  \\ 
<scenario> \\ 
Climate Analysis - Monitoring Urban Heat Trends   \\
</scenario>   \\

\#\# Generated Contexts:  \\
<context>  \\
temperatures: Daily recorded temperatures in a city, analyzed for urban heat trends.   \\
threshold: Critical temperature level indicating hazardous or abnormal heat.  \\
</context>   \\

\end{tcolorbox}

\subsection{Example for Context Generator Agent}
\begin{tcolorbox}[
    colback=orange!10!white,        
    colframe=orange,               
    title=Example for Context Generator Agent,
    fonttitle=\bfseries,
    coltitle=white,              
    colbacktitle=orange,          
    boxrule=0.8pt,               
    arc=1mm,                     
    enhanced,
    fontupper=\small
]
\ttfamily 
Given the natural language problem description, input types, and a real-world scenario. For each variable in the problem, provide a meaningful context tailored to the given scenario. The context should explain how each variable is involved in or relates to the scenario, ensuring practical relevance. \\ 

Please ensure for to put your meaningful context in <context></context> tags. \\

\# Problem Description: \\
$<$problem\_description$>$  \\
\textcolor{pyblue}{You're given a list of deposit and withdrawal operations on a bank account that starts with zero balance. Your task is to detect if at any point the balance of account fallls below zero, and at that point function should return True. Otherwise it should return False.} \\
$<$/problem\_description$>$ \\

\# Input Types: \\ 
$<$input\_types$>$  \\
\textcolor{pyred}{operations: list of int}   \\ 
$<$/input\_types$>$  \\

\# Real-world Scenario: \\
\textcolor{pygreen}{Education - Adaptive Learning Assessment and Skill Gap Analysis}        \\
$<$/scenario$>$          \\

\# Instructions:          \\
- For each variable in the input types, generate only one context that highlights its role or significance within the problem and scenario. \\
- The context should help to clarify the variable’s meaning and importance, ensuring that it fits into the given real-world scenario. \\ 
- Provide only the contexts for the variables (no additional reasoning steps).  \\

\# Example:  \\
\#\# Problem Description Example:  \\ 
$<$problem\_description$>$         \\
Determine if the average temperature in a city exceeds a certain threshold during a week.  \\
$<$/problem\_description$>$         \\

\#\# Input Types:         \\
$<$input\_types$>$  \\
temperatures: list of float  \\
threshold: float   \\
$<$/input\_types$>$  \\

\#\# Scenario:  \\ 
<scenario> \\ 
Climate Analysis - Monitoring Urban Heat Trends   \\
</scenario>   \\

\#\# Generated Contexts:  \\
<context>  \\
temperatures: Daily recorded temperatures in a city, analyzed for urban heat trends.   \\
threshold: Critical temperature level indicating hazardous or abnormal heat.  \\
</context>   \\

\end{tcolorbox}

\subsection{Prompt for Prompt Rewriter Agent}
\begin{tcolorbox}[
    colback=gray!10!white,        
    colframe=black,               
    title=Prompt for Prompt Rewriter Agent,
    fonttitle=\bfseries,
    coltitle=white,              
    colbacktitle=black,          
    boxrule=0.8pt,               
    arc=1mm,                     
    enhanced                    
]
\ttfamily 
Given a seed programming problem description, a selected real-world scenario, and contextualized input variables, rewrite the original problem to make it relevant to the scenario. The rewritten problem should: \newline
- Preserve the original problem's complexity and constraints. \\
- Ensure the problem remains solvable with the same solution approach. \\
- Be clear, concise, and logically consistent within the new context. \\
- Just return the rewritten problem description without any additional commentary or steps, and do not include any input output demons in your problem description. \\ 
- Limit the new rewritten problem description to 1-3 sentences. \\
- Make sure your rewritten problem description is clear, concise and contains no unnecessary information. \\

Please ensure to put your rewritten problem description in <new\_problem></new\_problem> tags. \\

\# Problem Description: \\
\textbf{\{PROBLEM DESCRIPTION\}} \\ 

\# Real-World Scenario: \\
\textbf{\{SCENARIO\}}  \newline

\# Contextualized Input Variables: \\
\textbf{\{INPUT VARIABLES\}}  \newline
\end{tcolorbox}

\subsection{Example for Prompt Rewriter Agent}
\begin{tcolorbox}[
    colback=orange!10!white,        
    colframe=orange,               
    title=Example for Prompt Rewriter Agent,
    fonttitle=\bfseries,
    coltitle=white,              
    colbacktitle=orange,          
    boxrule=0.8pt,               
    arc=1mm,                     
    enhanced                    
]
\ttfamily 
Given a seed programming problem description, a selected real-world scenario, and contextualized input variables, rewrite the original problem to make it relevant to the scenario. The rewritten problem should: \newline
- Preserve the original problem's complexity and constraints. \\
- Ensure the problem remains solvable with the same solution approach. \\
- Be clear, concise, and logically consistent within the new context. \\
- Just return the rewritten problem description without any additional commentary or steps, and do not include any input output demons in your problem description. \\ 
- Limit the new rewritten problem description to 1-3 sentences. \\
- Make sure your rewritten problem description is clear, concise and contains no unnecessary information. \\

Please ensure to put your rewritten problem description in <new\_problem></new\_problem> tags. \\

\# Problem Description: \\
\textcolor{pyblue}{You're given a list of deposit and withdrawal operations on a bank account that starts with zero balance. Your task is to detect if at any point the balance of account fallls below zero, and at that point function should return True. Otherwise it should return False.} \\ 

\# Real-World Scenario: \\
\textcolor{pyred}{Social Networking - Advanced Content Recommendation and User Interest Matching}
 \newline

\# Contextualized Input Variables: \\
\textcolor{pygreen}{A sequence of user interaction events (deposits/withdrawals) representing content engagement metrics in a social networking platform, where each operation tracks how users interact with recommended content, potentially influencing their future content visibility and recommendation algorithm.}  \newline
\end{tcolorbox}







\subsection{Prompt for Validation Agent 1}
\begin{tcolorbox}[
    colback=gray!10!white,        
    colframe=black,               
    title=Prompt for Validation Agent 1,
    fonttitle=\bfseries,
    coltitle=white,              
    colbacktitle=black,          
    boxrule=0.8pt,               
    arc=1mm,                     
    enhanced                    
]
\ttfamily 
Assess whether the two given natural language instructions convey the same meaning. Respond with 'Yes' if they do, or 'No' if they do not.\\

Please ensure your answer is either "Yes" or "No". \\

\# Instruction A: \\
\textbf{\{INSTRUCTION A\}} \\

\# Instruction B: \\
\textbf{\{INSTRUCTION B\}} \\

\end{tcolorbox}

\subsection{Example for Validation Agent 1}
\begin{tcolorbox}[
    colback=orange!10!white,        
    colframe=orange,               
    title=Example for Validation Agent 1,
    fonttitle=\bfseries,
    coltitle=white,              
    colbacktitle=orange,          
    boxrule=0.8pt,               
    arc=1mm,                     
    enhanced                    
]
\ttfamily 
Assess whether the two given natural language instructions convey the same meaning. Respond with 'Yes' if they do, or 'No' if they do not.\\

Please ensure your answer is either "Yes" or "No". \\

\# Instruction A: \\
\textcolor{pyblue}{For a given list of integers, return a tuple consisting of a sum and a product of all the integers in a list. Empty sum should be equal to 0 and empty product should be equal to 1.} \\

\# Instruction B: \\
\textcolor{pyred}{In an early disease risk prediction model, develop a function that processes a list of patient health metrics to calculate comprehensive risk assessment parameters. The function should compute two key aggregate indicators: the total sum of the patient's health metrics and the cumulative product of these metrics. For scenarios with no available health metrics, the sum should default to 0 and the product should default to 1, ensuring the model can handle incomplete patient data sets.}\\

\end{tcolorbox}







\subsection{Prompt for Validation Agent 2}
\begin{tcolorbox}[
    colback=gray!10!white,        
    colframe=black,               
    title=Prompt for Validation Agent 2,
    fonttitle=\bfseries,
    coltitle=white,              
    colbacktitle=black,          
    boxrule=0.8pt,               
    arc=1mm,                     
    enhanced                    
]
\ttfamily 
Does the following code solve the problem described in the Instruction? Provide your answer as either 'Yes' or 'No' only. \\

\# Instruction: \\
\textbf{\{INSTRUCTION\}} \\

\# Code Solution: \\
\textbf{\{CODE SOLUTION\}} \\
\end{tcolorbox}

\lstdefinestyle{mypython}{
  language=Python,
  basicstyle=\ttfamily\small,
  keywordstyle=\color{blue},
  stringstyle=\color{red!60!brown},
  commentstyle=\color{gray},
  showstringspaces=false,
  breaklines=true,
  frame=none
}


\newtcblisting{pythonbox}{
  colback=orange!10!white,
  colframe=orange,
  title=,
  coltitle=white,
  colbacktitle=orange,
  boxrule=0.0pt,
  arc=1mm,
  enhanced,
  listing only,
  listing options={
    style=mypython,
    xleftmargin=0pt,    
    aboveskip=0pt,      
    belowskip=0pt
  }
}
  
\subsection{Example for Validation Agent 2}
\begin{tcolorbox}[
    colback=orange!10!white,        
    colframe=orange,               
    title=Example for Validation Agent 2,
    fonttitle=\bfseries,
    coltitle=white,              
    colbacktitle=orange,          
    boxrule=0.8pt,               
    arc=1mm,                     
    enhanced                    
]
\ttfamily 
Does the following code solve the problem described in the Instruction? Provide your answer as either 'Yes' or 'No' only. \\

\# Instruction: \\
\textcolor{pyblue}{In a bank's loan risk assessment process, analyze a list of an applicant's key financial metrics to compute an aggregate financial risk score. Calculate the total sum of these financial indicators and their cumulative product to provide a comprehensive risk evaluation metric. For applicants with no financial history, the sum should default to 0 and the product should default to 1, ensuring a standardized risk assessment approach even for new customers.}\\

\# Code Solution: \\
\begin{pythonbox}
from typing import List, Tuple

def sum_product(numbers: List[int]) -> Tuple[int, int]:
    sum_value = 0
    prod_value = 1
    for n in numbers:
        sum_value += n
        prod_value *= n
    return sum_value, prod_value
\end{pythonbox}
\end{tcolorbox}
\section{Human Verification}
\label{appendix:human_verification}

To add an additional layer of validation between the original and \texttt{DyCodeEval}-generated prompts, we perform a small-scale manual verification. Given a benchmark dataset and the corresponding generated questions, we randomly sample $N = 30$ problem pairs from each dataset (60 in total), where each pair consists of a benchmark problem and its generated variant.
Each pair is independently reviewed by two graduate-level students to assess whether the core algorithm and complexity are preserved. In cases of disagreement, the reviewers discuss the discrepancies until consensus is reached. Out of the 60 reviewed pairs, the annotators initially disagreed on three but were able to resolve all disagreements through discussion, resulting in an overall agreement rate of 95\%.

\end{document}